\documentstyle[twoside,epsfig]{article}

\catcode`\@=11
\long\def\@makefntext#1{
\protect\noindent \hbox to 3.2pt {\hskip-.9pt
$^{{\eightrm\@thefnmark}}$\hfil}#1\hfill}		

\def\thefootnote{\fnsymbol{footnote}}
\def\@makefnmark{\hbox to 0pt{$^{\@thefnmark}$\hss}}	

\def\ps@myheadings{\let\@mkboth\@gobbletwo
\def\@oddhead{\hbox{}
\rightmark\hfil\eightrm\thepage}
\def\@oddfoot{}\def\@evenhead{\eightrm\thepage\hfil
\leftmark\hbox{}}\def\@evenfoot{}
\def\sectionmark##1{}\def\subsectionmark##1{}}

\oddsidemargin=\evensidemargin
\renewcommand{\thefootnote}{\fnsymbol{footnote}}

\newcounter{sectionc}
\newcounter{subsectionc}
\newcounter{subsubsectionc}
\renewcommand{\section}[1] {\vspace{12pt}\addtocounter{sectionc}{1}
\setcounter{subsectionc}{0}\setcounter{subsubsectionc}{0}\noindent
	{\tenbf\thesectionc. #1}\par\vspace{5pt}}
\renewcommand{\subsection}[1] {\vspace{12pt}
\addtocounter{subsectionc}{1}\setcounter{subsubsectionc}{0}\noindent
	{\bf\thesectionc.\thesubsectionc.
        {\kern1pt \bfit #1}}\par\vspace{5pt}}
\renewcommand{\subsubsection}[1] {\vspace{12pt}
\addtocounter{subsubsectionc}{1}\noindent
        {\tenrm\thesectionc.\thesubsectionc.\thesubsubsectionc.
	{\kern1pt \tenit #1}}\par\vspace{5pt}}

\newcounter{appendixc}
\newcounter{subappendixc}[appendixc]
\newcounter{subsubappendixc}[subappendixc]
\renewcommand{\thesubappendixc}{\Alph{appendixc}.
        \arabic{subappendixc}}
\renewcommand{\thesubsubappendixc}{\Alph{appendixc}.
        \arabic{subappendixc}.\arabic{subsubappendixc}}

\renewcommand{\appendix}[1] {\vspace{12pt}
        \refstepcounter{appendixc}
        \setcounter{figure}{0}
        \setcounter{table}{0}
        \setcounter{lemma}{0}
        \setcounter{theorem}{0}
        \setcounter{corollary}{0}
        \setcounter{definition}{0}
        \setcounter{equation}{0}
        \renewcommand{\thefigure}{\Alph{appendixc}.\arabic{figure}}
        \renewcommand{\thetable}{\Alph{appendixc}.\arabic{table}}
        \renewcommand{\theappendixc}{\Alph{appendixc}}
        \renewcommand{\thelemma}{\Alph{appendixc}.\arabic{lemma}}
        \renewcommand{\thetheorem}{\Alph{appendixc}.\arabic{theorem}}
        \renewcommand{\thedefinition}{\Alph{appendixc}.
         \arabic{definition}}
        \renewcommand{\thecorollary}{\Alph{appendixc}.
         \arabic{corollary}}
        \renewcommand{\theequation}{\Alph{appendixc}.
         \arabic{equation}}
        \noindent{\tenbf Appendix \theappendixc #1}\par\vspace{5pt}}
\newcommand{\subappendix}[1] {\vspace{12pt}
        \refstepcounter{subappendixc}
        \noindent{\bf Appendix \thesubappendixc. {\kern1pt \bfit #1}}
	\par\vspace{5pt}}
\newcommand{\subsubappendix}[1] {\vspace{12pt}
        \refstepcounter{subsubappendixc}
        \noindent{\rm Appendix \thesubsubappendixc.
        {\kern1pt \tenit #1}}\par\vspace{5pt}}

\newcommand{\textlineskip}{\baselineskip=13pt}
\newcommand{\smalllineskip}{\baselineskip=10pt}

\def\eightcirc{
\begin{picture}(0,0)
\put(4.4,1.8){\circle{6.5}}
\end{picture}}
\def\eightcopyright{\eightcirc\kern2.7pt\hbox{\eightrm c}}

\newcommand{\pub}[1]{{\begin{center}\footnotesize\smalllineskip
	Preprint No. #1\\
	\end{center}
	}}

\def\abstracts#1#2#3{{
	\centering{\begin{minipage}{4.5in}\baselineskip=10pt
        \footnotesize
	\parindent=0pt #1\par
	\parindent=15pt #2\par
	\parindent=15pt #3
	\end{minipage}}\par}}

\renewenvironment{thebibliography}[1]
	{\frenchspacing
	 \ninerm\baselineskip=11pt
	 \begin{list}{\arabic{enumi}.}
	{\usecounter{enumi}\setlength{\parsep}{0pt}
	 \setlength{\leftmargin 12.7pt}{\rightmargin 0pt}
	 \setlength{\itemsep}{0pt} \settowidth
	{\labelwidth}{#1.}\sloppy}}{\end{list}}

\newcounter{itemlistc}
\newcounter{romanlistc}
\newcounter{alphlistc}
\newcounter{arabiclistc}

\newcommand{\fcaption}[1]{
        \refstepcounter{figure}
        \setbox\@tempboxa = \hbox{\footnotesize Fig.~\thefigure. #1}
        \ifdim \wd\@tempboxa > 5in
           {\begin{center}
        \parbox{5in}{\footnotesize\smalllineskip Fig.~\thefigure. #1}
            \end{center}}
        \else
             {\begin{center}
             {\footnotesize Fig.~\thefigure. #1}
              \end{center}}
        \fi}

\newcommand{\tcaption}[1]{
        \refstepcounter{table}
        \setbox\@tempboxa = \hbox{\footnotesize Table~\thetable. #1}
        \ifdim \wd\@tempboxa > 5in
           {\begin{center}
        \parbox{5in}{\footnotesize\smalllineskip Table~\thetable. #1}
            \end{center}}
        \else
             {\begin{center}
             {\footnotesize Table~\thetable. #1}
              \end{center}}
        \fi}

\def\@citex[#1]#2{\if@filesw\immediate\write\@auxout
	{\string\citation{#2}}\fi
\def\@citea{}\@cite{\@for\@citeb:=#2\do
	{\@citea\def\@citea{,}\@ifundefined
	{b@\@citeb}{{\bf ?}\@warning
	{Citation `\@citeb' on page \thepage \space undefined}}
	{\csname b@\@citeb\endcsname}}}{#1}}

\newif\if@cghi
\def\cite{\@cghitrue\@ifnextchar [{\@tempswatrue
	\@citex}{\@tempswafalse\@citex[]}}
\def\citelow{\@cghifalse\@ifnextchar [{\@tempswatrue
	\@citex}{\@tempswafalse\@citex[]}}
\def\@cite#1#2{{$\null^{#1}$\if@tempswa\typeout
	{IJCGA warning: optional citation argument
	ignored: `#2'} \fi}}

\def\pmb#1{\setbox0=\hbox{#1}
	\kern-.025em\copy0\kern-\wd0
	\kern.05em\copy0\kern-\wd0
	\kern-.025em\raise.0433em\box0}

\def\fnt#1#2{\footnotetext{\kern-.3em
	{$^{\mbox{\scriptsize #1}}$}{#2}}}

\def\fpage#1{\begingroup
\voffset=.3in
\thispagestyle{empty}\begin{table}[b]\centerline{\footnotesize #1}
	\end{table}\endgroup}

\font\tenrm=cmr10
\font\tenit=cmti10
\font\tenbf=cmbx10
\font\bfit=cmbxti10 at 10pt
\font\ninerm=cmr9

\font\eightrm=cmr8

\def\qed{\hbox{${\vcenter{\vbox{		
   \hrule height 0.4pt\hbox{\vrule width 0.4pt height 6pt
   \kern5pt\vrule width 0.4pt}\hrule height 0.4pt}}}$}}

\renewcommand{\thefootnote}{\fnsymbol{footnote}}

\input epsf
\global\arraycolsep=2pt
\def\spose#1{\hbox to 0pt{#1\hss}}
\def\lsim{\mathrel{\spose{\lower 3pt\hbox{$\mathchar"218$}}
 \raise 2.0pt\hbox{$\mathchar"13C$}}}
\def\gsim{\mathrel{\spose{\lower 3pt\hbox{$\mathchar"218$}}
 \raise 2.0pt\hbox{$\mathchar"13E$}}}

\renewcommand{\theequation}{\thesection.\arabic{equation}}
\def\laq{\raise 0.4ex\hbox{$<$}\kern -0.8em\lower 0.62
ex\hbox{$\sim$}}
\def\gaq{\raise 0.4ex\hbox{$>$}\kern -0.7em\lower 0.62
ex\hbox{$\sim$}}

\def\beq{\begin{equation}}
\def\eeq{\end{equation}}
\def\bea{\begin{eqnarray}}
\def\eea{\end{eqnarray}}

\def \ra {\rightarrow}

\def \la {\lambda}

\def \Da {\Delta}

\def \a {\alpha}
\def \ap {\alpha^{\prime}}

\def \ga {\gamma}

\def \da {\delta}

\def \r {\rho}
\def \om {\omega}
\def \Om {\Omega}
\def \noi {\noindent}

\begin{document}

\begin{titlepage}

\begin{flushright}
DFTT-43/97\\
gr-qc/9707034
\end{flushright}

\vspace{3 cm}

\begin{center}
\Large\bf Testing String Cosmology\\
with Gravity Wave Detectors
\end{center}

\vspace{2cm}

\begin{center}
M. Gasperini\\
{\sl Dipartimento di Fisica Teorica, Universit\`a di Torino,}\\
{\sl Via P. Giuria 1, 10125 Turin, Italy}\\
and\\
{\sl Istituto Nazionale di Fisica Nucleare, Sezione di Torino, Turin, Italy}\\
\end{center}

\vspace{2cm}

\begin{abstract}
\noi
The general properties of the gravity wave backgrounds of
cosmological origin are reviewed and briefly discussed, with emphasis
on the relic background expected from an early pre-big bang phase
typical of string cosmology models.

\end{abstract}

\vspace{2cm}
\begin{center}
------------------------------

\vspace{2cm}
To appear in \\
{\sl Proc. of the ``Second Edoardo Amaldi Conference on
Gravitational Waves"}\\ 
CERN, July 1997 -- 
Eds. E. Coccia et al. \\
(World Scientific, Singapore)
\end{center}
 \vspace{1.5cm}
\vfill

\end{titlepage}


\normalsize\textlineskip
\thispagestyle{empty}
\setcounter{page}{1}


\vspace*{0.11truein}

\fpage{1}

\centerline{\bf TESTING STRING COSMOLOGY}
\centerline{\bf WITH GRAVITY WAVE DETECTORS} 
\vspace*{0.27truein}

\centerline{\footnotesize MAURIZIO GASPERINI}
\vspace*{0.015truein}
\centerline{\footnotesize\it Dipartimento di Fisica Teorica, 
Universit\`a di Torino,}
\baselineskip=10pt
\centerline{\footnotesize  {\it Via P. Giuria 1, 10125, Turin, Italy}}
\baselineskip=10pt
\centerline{\footnotesize and {\it Istituto Nazionale di Fisica Nucleare,
Sezione di Torino, Turin, Italy}}

\vspace*{0.3truein}
\abstracts
{The general properties of the gravity wave backgrounds of
cosmological origin are reviewed and briefly discussed, with emphasis
on the relic background expected from an early pre-big bang phase
typical of string cosmology models. }
{}{}
\vspace*{0.225truein}
\pub{DFTT-43/97;~~~~~~ E-print Archives: gr-qc/9707034}
\vspace*{0.8pt}\textlineskip

\textheight=7.8truein
\setcounter{footnote}{0}
\renewcommand{\thefootnote}{\alph{footnote}}

\vspace*{0.125truein}

\renewcommand{\theequation}{1.\arabic{equation}}
\setcounter{equation}{0}
\section{Introduction}
\label{sec:1}
\noindent
The amplification of the quantum fluctuation of the vacuum, and the
generation of primordial perturbation spectra, is one of the most
celebrated aspects of the standard inflationary scenario\cite{1}. In
particular, the amplification of the transverse and trace-free part of
the metric fluctuations (which are decoupled from the sources, in the
linear approximation), leads to the formation of a primordial gravity
wave background that may survive, nearly unchanged, down to the
present time\cite{2}. Such background is characterized by three main
properties: 
\begin{itemize}
\item{}it is stochastic, because of its quantum origin;
\item{}the present fluctuation amplitude, for each Fourier mode, is
directly related to the curvature scale of the Universe at the time of 
first horizon crossing of that mode;
\item{}the spectral distribution of the amplitude is determined by the
kinematic behavior of the scale factor at the horizon crossing epoch.
\end{itemize}
It is thus obvious that a cosmic background of relic gravity waves
retains the imprint of the primordial dynamics, and may provide
direct information on the very early history of our Universe\cite{3}.

Unfortunately, however, the relic background expected in the
frequency band of the present interferometric and resonant-mass
detectors, according to the standard inflationary scenario, is by far too
low to be detected\cite{4}, both in first and second generation
experiments. The reason for this disappointing conclusion is that the
background, characterized by a flat or decreasing spectrum, is strongly
constrained by the large scale anisotropy observed by COBE at the
present horizon scale\cite{5}, $\Da T/T\sim 10^{-5}$. The energy
density $\Om_G$ of the graviton background, in critical units, is thus
bounded at high frequency by the condition\cite{6}:
\beq
\Om_G~\laq ~\Om_{CMB}\left(\Da T \over T\right)_{COBE}^2~\sim  
~10^{-14}, \,\,\,\,\,\,\, \,\,\,\,\,\,\om ~\gaq ~10^{-16}~{\rm Hz} . 
\label{11}
\eeq
where $\Om_{CMB}$ is the present fraction of critical energy density in
the form of Cosmic Microwave Background (CMB) radiation. The above
bound is saturated by a flat, Harrison-Zeldovich spectral distribution
(assuming that the spectrum is normalized at the horizon scale by
the COBE data), while decreasing spectra always lead to a lower
$\Om_G$. For comparison, the maximal sensitivity to a stochastic
background\cite{7} expected in the context of the Advanced LIGO
project is only $\Om_G \sim 
10^{-10}$, at a frequency $\nu \sim 10^{2}$ Hertz. 

The situation is instead more rosy for the pre-big bang models\cite{8}
formulated in the context of string cosmology. In that case, the
spectral energy density of the background tends to grow very fast
with frequency, and it is thus too low, at the COBE scale, to be
constrained by the observed CMB anisotropy\cite{8,9} (the constraint
from pulsar timing data\cite{10}, $\Om_G~\laq~ 10^{-8}$ at a frequency 
$\nu \sim 10^{-8}$ Hz, 
can also be easily satisfied). At high frequency, the
maximal intensity of the background is simply controlled by the
fundamental ratio between string ($M_s$) and Planck ($M_P$) mass,
which sets the natural value of the final inflation scale, and which is
expected\cite{11} to be a number in the range $0.3-0.03$. One thus
obtain the bound\cite{12,12a}
\beq
\Om_G~ \laq ~ \Om_{CMB}\left(M_s \over M_P\right)^2~\laq 
~10^{-5} ,
\label{12}
\eeq
which corresponds to a possible enhancement of nine orders of
magnitude, at high frequency, 
with respect to the peak intensity (\ref{11}) typical of the
standard inflationary scenario. 

The amplification of the vacuum fluctuations, however, is not the only
mechanism leading to the formation of 
a primordial gravity wave background.
There are processes, in the context of the standard inflationary models,
producing backgrounds which are mainly localized at high
frequency, and which may thus evade the constraint (\ref{11}). Three
possible backgrounds, in particular, should be mentioned. The
background due to gravitational radiation from cosmic
strings\cite{13} and other topological defects\cite{14}, the
background due to bubble collision at the end of a first order
phase transition\cite{15}, in extended inflation models, and the
background produced by parametric resonance effects\cite{16},
during the so-called ``preheating" phase.

A global view of these possible primordial relic backgrounds, in the
frequency range $\om > 1$ Hz, is qualitatively sketched in Fig. 1, where
I have plotted the present value of the spectral energy density (in
critical units) of the background:
\bea
&&
\Om_G(\om, t_0)={\om\over \r_c(t_0)}{ d\r_G(\om, t_0)\over d\om},
~~~~~~~~\r_c(t_0)={3M_p^2H_0^2\over 8\pi},   \nonumber\\ 
&&
H_0=h_{100}\times (100~ {\rm km ~sec^{-1} Mpc^{-1}}) .
\label{13}
\eea
The bold solid line of Fig. 1 define the allowed region for a background
produced through the parametric amplification of the vacuum
fluctuations, in string cosmology (upper lines) and in standard
inflationary cosmology (lower lines). The dashed lines represent
possible spectra for backgrounds obtained from topological defects,
phase transitions and resonant inflaton oscillations. In the first case
the plotted spectrum refers to the maximal allowed background
associated to graviton radiation from cosmic strings\cite{13}. In the
other two cases the spectrum is strongly dependent on the final
reheating temperature, $T_r$. The example of phase transitions 
illustrated in Fig. 1 refers to $T_r\sim 10^8 - 10^9$ GeV, but higher
backgrounds are possible\cite{15} for higher values of $T_r$. 

\begin{figure}[htb]
\vspace{10cm}
\includegraphics{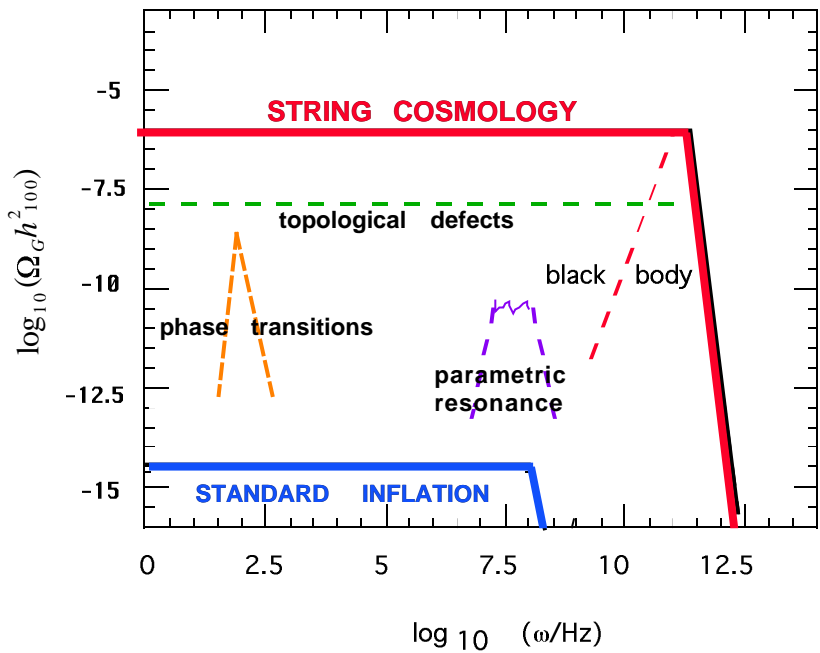}
{\caption{\label{fig:f1}
{\sl Possible gravity wave backgrounds of cosmological origin (dashed
lines), in the frequency range $\om > 1$ Hz. The bold solid lines define
the allowed region for a background obtained from the quantum
fluctuations of the metric, in string cosmology (upper lines) and in
standard inflationary cosmology (lower lines).}}}
\end{figure}

Also shown in Fig. 1 is a thermal black-body spectrum
corresponding to a temperature $T_0\sim 1^0K$. In the
standard scenario a thermal gravity wave background might
originate at the Planck scale, when the temperature is high enough
to maintain gravitons in thermal equilibrium. However,
such a background  should be
strongly diluted (with respect to the present CMB radiation) by the
action of the subsequent inflationary phase, occurring at curvature
scales lower than Planckian. As a consequence, the surviving 
spectrum should correspond today to an effective temperature so
depressed to be practically invisible. In string cosmology, on the
contrary, a graviton background with the typical low-frequency
slope of a black-body spectrum, 
$\Om_G \sim \om^3$, is possibly generated by the sudden
transition from an initial dilaton-driven phase to the standard
radiation-dominated era\cite{8,12}. Modulo logarithmic corrections,
such a spectrum may simulate a relic thermal background of
Planckian origin\cite{17,17a}, with a typical effective temperature
which is just of the same order as that of the thermal spectrum
shown in Fig. 1. 

The rest of this paper will be devoted to explain, and discuss, the big
difference between the two allowed regions relative to a vacuum
fluctuation spectrum. Before starting the discussion, however, it may
be be appropriate to recall that at present 
the best experimental upper
bound on a possible stochastic background, in the frequency range of
Fig. 1, is provided by the cross-correlation of the two 
resonant-mass detectors  NAUTILUS and EXPLORER. The most recent
data imply\cite{19a}
\beq
\Om_G h_{100} ~ \laq ~60, ~~~~~~~~~~~~~~~~~~~
\nu\simeq 907 ~{\rm Hz}.
\label{14}
\eeq
improving by about an order of magnitude the previous upper
limit\cite{18} obtained with EXPLORER.  
A much better sensitivity, $\Om_G h_{100}\sim 10^{-3}-10^{-5}$, is
expected to be reached in the near future by the cross-correlation of
NAUTILUS, EXPLORER and AURIGA\cite{18,19}, and by the first operating
version\cite{7,21a} of  LIGO and VIRGO, in the frequency bands $\nu
\sim 10^{3}$ Hz and $\nu \sim 10^{2}$ Hz, respectively. 
Similar sensitivities are expected from the cross-correlation of a bar
and an interferometer\cite{21b}. These
sensitivities are still outside, but not so far off, the upper border of
the allowed region in Fig. 1. To get inside we have to wait, for
instance, for the cross-correlation of two spherical resonant-mass
detectors\cite{19,21c},  with expected sensitivity $\Om_G \sim
10^{-7}$ around  $\nu \sim 10^{3}$ Hz,   or for the advanced version of
the interferometric detectors\cite{7,21a},
with expected sensitivity $\Om_G \sim 10^{-10}$ around  
$\nu \sim 10^{2}$ Hz. In both cases the detectors will cross the border
of the allowed region, and will explore, for the first time, the
parameter space of string cosmology and of Planck scale physics.  

\vskip 1 cm
\renewcommand{\theequation}{2.\arabic{equation}}
\setcounter{equation}{0}
\section{Properties of the string cosmology background}
\label{sec:2}
\noindent
In string cosmology, like in the standard inflationary scenario, the
generation of a gravity wave background from the ground state
configuration is due to a process of parametric amplification of the
metric fluctuations, under the action of the
cosmological gravitational field playing the role of the external
``pumping" force\cite{2}. The basic difference from the standard
scenario arises from an enhancement of this amplification process in the
high frequency sector. This enhancement can be ascribed to three
independent mechanisms:
\begin{itemize}
\item{}the growth of the curvature during the phase of accelerated
evolution, and the consequent growth with frequency of the
perturbation spectrum;
\item{} the possible growth in time of the comoving amplitude of 
perturbations even outside the horizon, instead of its ``freezing"
typical of standard inflation;
\item{}the additional amplification due to the higher-derivative
terms that must be added to the effective action when the curvature
becomes large in string units.
\end{itemize}
\bigskip

The first two effects are a consequence of the special kinematic of
the phase of accelerated pre-big bang evolution, characterized by
shrinking event horizons\cite{8}. During such a phase the scale factor
can be parametrized, in conformal time $\eta$ and in the Einstein frame,
as  
\beq
a(\eta)= (-\eta)^\a, ~~~~~~~~~~~ \eta < 0, ~~~~~~~~~~~
\a \geq -1.
\label{21}
\eeq
The tensor perturbation equation\cite{2}, for the Fourier component of
each polarization mode of comoving amplitude $h_k$,
\beq
\psi_k''+\left(k^2- {a''\over a}\right) \psi_k =0 , ~~~~~~~~~~~~
\psi_k=a h_k,
\label{22}
\eeq
outside the horizon ($|k\eta|\ra 0$) has the general asymptotic
solution: 
\beq
h_k= A_k +B_k \left|\eta\right|^{1-2\a} , ~~~~~~~~~~~~~~~
\eta \ra 0_-
\label{23}
\eeq
($A_k, B_k$ are integration constant, and the prime denotes
differentiation with respect to $\eta$). For the metric (\ref{21}) we
may thus distinguish two possibilities.

If $\a <1/2$, $h_k$ tends to stay constant outside the horizon. By
normalizing the canonical variable $\psi_k$ to a vacuum fluctuation
spectrum, at the time of horizon crossing $|\eta|=k^{-1}$, we obtain
asymptotically $\psi_k \sim (a/a_{hc})k^{-1/2}$, with consequent
spectral amplitude $k^{3/2}|h_k|\sim |a\eta|^{-1}_{hc} \sim |H|_{hc}$,
where $H=\dot a /a =a'/a^2$ (a dot denotes differentiation with
respect to the cosmic time $t$, defined by $dt= ad\eta$). This amplitude
grows with $k$ because higher frequency modes cross the horizon
later in time, and then at higher values of $|H|$, since $|H|$ is growing
for the metric (\ref{21}). 

If $\a >1/2$ there is an additional growth in time of $h_k$ itself
outside the horizon, according to the asymptotic solution (\ref{23}).
This second effect is usually excluded in the standard inflationary
models, characterized by $\a <0$. In string cosmology this effect is due
to the accelerated growth of the dilaton\cite{9,20}, which accompanies
the growth of the curvature scale, and which transforms the scale
factor kinematic of the Einstein frame (where the dilaton is decoupled
from tensor perturbations) into a fast, accelerated contraction\cite{8},
with $\a >0$. In the limiting case $\a=1/2$, corresponding to the
simplest 
low-energy gravi-dilaton model, the growth in time of $h_k$ is simply
logarithmic\cite{17}, $h_k \sim \ln |k\eta|$, and can be neglected for
an order of magnitude estimate of the spectrum.

A third, additional contribution to the amplification is due to the fact
that, because of the growth of the curvature during the pre-big bang
phase, the late-time evolution of perturbations 
takes place in the high-curvature regime. In this regime, the higher
derivative corrections predicted by string theory may become
important, and should be included into the effective action. To the first
non-leading order of the 
so-called $\ap$ expansion, where $\ap=\la_s^2$ is the
basic string length parameter, the corrected action can be
written as\cite{21}
\beq
S=-{1\over 2\la_s^{d-1}}\int d^{d+1}x \sqrt{|g|}e^{-\phi} \left[ R+
(\nabla \phi)^2-{k\ap \over 4} \left(R^2_{GB} - 
(\nabla \phi)^4\right)\right] ,
\label{24}
\eeq
where we have used a convenient field-redefinition that introduces
the Gauss-Bonnet invariant $R_{GB}$, thus eliminating higher-than-second
derivatives from the equations of motion. Such higher-curvature
corrections are important because they tend to stop the growth of
the curvature and of the dilaton\cite{22}, driving the Universe to a
phase with $H=$ const, $\dot\phi=$ const. 

The perturbation of the action (\ref{24}) around a homogeneous and
isotropic background solution, $a(\eta), \phi(\eta)$, leads to a 
generalized tensor perturbation equation\cite{23}: 
\bea
&&
\psi_k^{\prime\prime}+\left[k^2-
{z^{''}\over z}+{k^2\over z^2}(y^2-z^2)\right]\psi_k=0,
\,\,\,\,\,\,\psi_k =z h_k,\nonumber \\ 
&&
z^2(\eta)=e^{-\phi}\left(a^2-\ap {a'\over a}\phi'\right), \,\,\,\,\,
y^2(\eta)=e^{-\phi}\left[a^2+\ap 
\left(\phi^{\prime 2}-\phi^{''}+
 {a'\over a}\phi'\right)\right]
\label{25}
\eea
(here $a$ is the metric scale factor  in the frame of the action
(\ref{24})). This equation includes the high-curvature corrections to
first order in $\ap$, and reduces to the low-energy equation 
in the limit $\ap \ra 0$. 

The results of a numerical integration of eq. (\ref{25}) are illustrated
in Fig. 2, where we show the evolution in cosmic time of the 
comoving perturbation amplitude $|h_k|$, computed with and
without the $\ap$ corrections\cite{23}. In both cases the amplitude
is oscillating inside the horizon, and frozen outside the horizon.
When the $\ap$ corrections are included, however, the final
asymptotic amplitude is enhanced with respect to the amplitude
obtained, for the {\em same} mode and in the {\em same}
background, without the $\ap$ corrections. 

\begin{figure}[htb]
\vspace{5cm}
\includegraphics{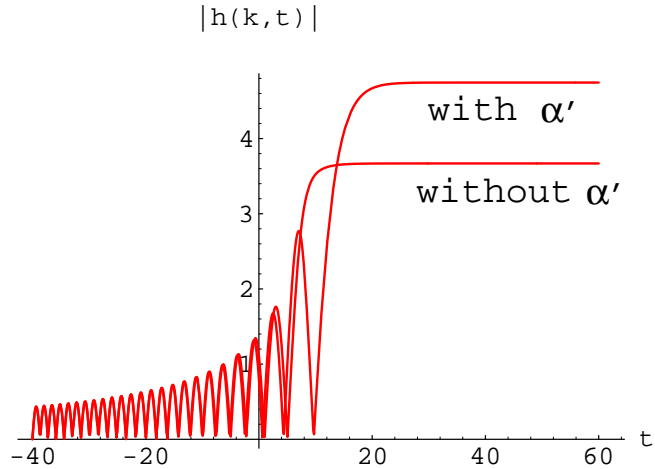}
{\caption{\label{fig:f2}
{\sl Time-evolution of the comoving amplitude $|h_k(t)|$, according to 
a numerical integration of eq. (\ref{25}) with and without the
high-curvature corrections.}}}
\end{figure}

This enhancement is the same for all modes, and thus does not
affect the gravity wave spectrum computed with the low-energy
perturbation equation (\ref{22}). The effect of the high-curvature
corrections amounts to an overall rescaling, by a numerical factor of
the order of unity, of the total energy density of the background,
and may thus be neglected for an order of magnitude estimate of
graviton production.

By using eqs. (\ref{22}) and (\ref{25}) we can now predict some
general property of the gravity wave background expected in the
context of the pre-big bang scenario. Such predictions are to a large
extent model-independent, provided we accept that the Universe
becomes radiation-dominated soon after the end of the
high-curvature string phase. 

At low energy, i.e. for modes crossing the the horizon when the
$\ap$ corrections are still negligible, the slope of the spectrum can
be computed exactly, and for the metric (\ref{21}) with $\a=1/2$ we
find the nearly thermal behavior\cite{8,17}
\beq
\Om_G(\om) \propto \left(\om \over \om_s\right)^3\ln 
 \left(\om \over \om_s\right) , \,\,\,\,\,\,\,\,\,\,\,\,\,
\om < \om_s .
\label{26}
\eeq
Here $\om_s$ is the frequency scale at which high-curvature
string effects may become important. At high frequency the slope is
model-dependent, but in general flatter than cubic\cite{12,12a,24}
because the high-derivative\cite{22} 
and loop\cite{25} corrections tend to
stop the growth of the curvature and of the dilaton kinetic energy,
and thus tend to depress the slope of the metric perturbation
spectrum.

The maximal amplified frequency $\om_1$, i.e. the frequency
corresponding to the production of one graviton per polarization and
per unit phase space volume, can also be computed in a
model-independent way, an can be conveniently related to the
present CMB temperature $T_0$ as follows\cite{12a}:
\beq
\om_1(t_0) \simeq  T_0\left(M_s\over M_P\right)^{1/2}
\left(10^3\over n_r \right)^{1/12}\left(1-\da S\right)^{1/3} ,
\,\,\,\,\,
T_0 \simeq 3.6 \times 10^{11} {\rm Hz}
\label{27}
\eeq
Here $n_r \simeq 10^3$ is the total number of thermal degrees of
freedom in equilibrium at the beginning of the radiation era, and
$\da S$ is the fraction of present thermal entropy density due to all
reheating processes occurring well below the end of the string
phase. The occurrence of such processes would imply that the
radiation which becomes dominant at the end of the string phase is
only a fraction of the CMB radiation that we observe today, and this
would dilute the energy density of the gravity wave background
with respect to the present CMB energy density. 

The peak intensity
of the background, in this context, has to be of the same order as
the end-point energy density\cite{12a},
\beq
\Om_G(\om_1, t_0)= {\om_1^4(t_0)\over \pi^2\rho_c(t_0)}
\simeq 7\times 10^{-5}h_{100}^{-2}
\left(M_s\over M_P\right)^{2}
\left(10^3\over n_r \right)^{1/3}\left(1-\da S\right)^{4/3}.
\label{28}
\eeq
As a consequence, for $\om <\om_1$, the spectrum may be
decreasing or at most flat, leading to the allowed region illustrated
in Fig. 1 (where we have assumed $n_r=10^3$). 

The allowed region can be obviously extended also at frequencies
$\om < 1$ Hz. At lower frequencies, however, the upper border of
the region has to be slightly decreasing, in order to satisfy the
constraint coming from primordial nucleosynthesis\cite{26}, which
implies that the total integrated energy density of the background
cannot exceed, roughly, that of one massless degree of freedom in
thermal equilibrium. More precisely, nucleosynthesis implies the
bound\cite{12a}
\beq
h^2_{100}\int  \Om_G (\om, t_0) d \ln \om\, \laq \,
0.5\times 10^{-5} .
\label{29}
\eeq
This bound is compatible with, but almost completely saturated
(depending on $M_s/M_P$) by the peak intensity (\ref{28}). So, a
strictly flat maximal spectral density cannot be extended to
arbitrarily low frequencies ($\ll 1$ Hz)
without conflicting with the bound
(\ref{29}). At very low frequencies there are, in addition, stronger
phenomenological constraints from pulsar-timing data\cite{10} 
and COBE data\cite{5}, 
as discussed in the previous Section.

\vskip 1 cm
\renewcommand{\theequation}{3.\arabic{equation}}
\setcounter{equation}{0}
\section{Testing string cosmology models}
\label{sec:3}
\noindent
In the context of the pre-big bang scenario, any relic graviton
spectrum $\Om_G(\om)$ which reaches the end point with a slope
not larger than cubic (in the low frequency sector) 
is in principle allowed, like
the spectra represented by the bold solid lines of Fig. 3. Notice that
above the maximal frequency $\om_1$ the graviton production is
exponentially suppressed\cite{27}, and the spectrum must decrease
with the typical  high frequency 
behavior of a  Planckian distribution. 

For any given value of $n_r$ and $\da S$ there is a residual
uncertainty on the values $\om_1$ and $\Om_G(\om_1)$, according
to eqs. (\ref{27}) and (\ref{28}), corresponding to the present
theoretical uncertainty on the values of the fundamental string
theory parameter $M_s=\la_s^{-1}$. This uncertainty is
represented by the shaded boxes of Fig. 3, where we have chosen,
for illustrative purpose, $n_r=10^3$ and 
\beq
0.01 ~ \laq \,  \left(M_s\over M_P\right) \, \laq ~0.1 .
\label{31}
\eeq
To the left of the end point the spectrum can be at most flat, for the
simplest class of models considered in the previous section. In the
absence of significant reheating at scales much lower than the end
of the string phase, the maximal intensity of the gravity wave
background is thus expected within the dashed lines, in the band
labeled $\da S=0$. 

Also plotted in Fig. 3 is the corresponding band for the case that 99
per cent of the present large scale entropy is due to some
low-energy process occurring well below the string scale. Even in
that case, the expected peak intensity stays well above the full line
labeled ``de Sitter", and corresponding to the most optimistic
predictions of the standard inflationary scenario. 

\begin{figure}[htb]
\vspace{8cm}
\includegraphics{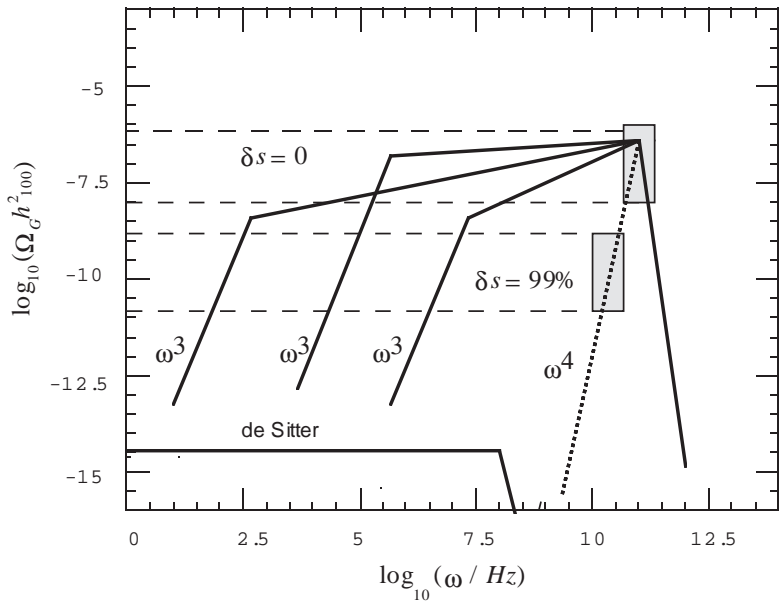}
{\caption{\label{fig:f3}
{\sl Possible allowed spectra for a typical class of string cosmology
models.}}}
\end{figure}

Entering the region where we may expect a signal
thus require a minimal sensitivity\cite{12,12a}
\beq
\Om_Gh_{100} ~\laq~10^{-6} , 
\label{32}
\eeq
or, in terms of the strain density $S_h(\nu)$,
\beq
S_h^{1/2}(\nu)~\laq ~3 \times 10^{-26}\left({\rm
kHz}\over \nu \right)^{3/2}~{\rm Hz}^{-1/2}, \,\,\,\,\,~~
S_h(\nu)={3 H_0^2 \over 4 \pi^2 \nu^3}\Om_G (\nu) .
\label{33}
\eeq
Any detector able to reach this limit will be already in a position to
receive a signal from the cosmic graviton background or, in case of
a negative result, to constrain the parameter space of the string
cosmology models\cite{12,17a}. In particular, any measure inside the
allowed region will provide significant information on the possible
value of the frequency $\om_s$ which marks the end of the 
low-energy
branch of the spectrum (\ref{26}), and on the corresponding 
value of the dilaton $\phi_s$ and of the
string coupling $g_s^2=\exp (\phi_s)$. 

The importance of determining the coordinates ($\om_s$ and
$\Om_G(\om_s)$) of this break point of the spectrum, signalling the
beginning of the high-curvature regime, is self-evident. From its
position in the plane of Fig. 3 we could immediately deduce the
duration in time ($\sim \om_s/\om_1$), and the rate of growth of
the curvature in Planck units ($\sim \Om_s/\Om_1$), for the 
``stringy" regime. This would impose important constraints on other
phenomenological aspects of the pre-big bang scenario indirectly
related to graviton production (such as the production of seeds for
the cosmic magnetic fields\cite{28}), as discussed
elsewhere\cite{17a,29}. 

Also, suppose to detect the high frequency part of the graviton
spectrum, namely a signal which grows with frequency, $\Om_G
\sim \om^\ga$, with a positive slope $\ga < 3$, as illustrated in Fig.
3. The intercept of that spectrum (extrapolated up to the GHz range)
with the ``one-graviton" line $\Om_G \sim \om_1^4$
(the dotted  line of Fig. 3), would give a first {\em
experimental} indication of the value of the fundamental ratio
$M_s/M_P$. 

Of course, the situation is not so simple. The possibility of detecting
a signal inside the allowed region depends on the shape of the
spectrum, and the shape, unlike the allowed region, is strongly
model-dependent. A detailed discussion of this point is outside the
scope of this paper; it will be enough to recall, in this context, that
there are two main classes of models which we may call\cite{29}
``minimal" and ``non-minimal", characterized by a different
evolution in time of the curvature scale and of the string coupling
$e^\phi$. In the minimal case the beginning of the radiation era
coincides with the end of the high-curvature string phase, in the
non-minimal case the coupling is still small at the end of the string
phase, and the radiation era begins much later. 

The main difference between the two cases is that in the second
case the effective potential which amplifies tensor perturbation,
according to eqs. (\ref{22}), (\ref{25}), is non-monotonic, and 
the highest frequency modes may reenter the horizon before
the beginning of the radiation era. This modifies the slope of the
spectrum in the high-frequency sector, with the possible
appearance of a negative power. The gravity wave spectrum may
thus become non-monotonic\cite{29}, and the peak may not coincide
any longer with the end point (see also [36] for a different
possibility of non-monotonic spectrum). 

These are good news from an experimental point of view, because
they make more probable a large detectable signal at frequencies
lower than the GHz band. However, they also provide a warning
against a too naive interpretation of possible future experimental
data, because of the complexity of the parameter space of the
string cosmology models.

\vskip 1 cm
\renewcommand{\theequation}{4.\arabic{equation}}
\setcounter{equation}{0}
\section{Conclusion}
\label{sec:4}
\noindent
Summarizing the results reported here, my conclusion is very
simple: there is no compelling reason (at present, and to the best of
my knowledge) to exclude the presence of a stochastic graviton
background of primordial origin, with an energy density as high as
\beq
\Om_g(\om)\sim 10^{-6}h_{100}^{-2},\,\,\,\,\,\,\,\,\,\,\,\,
1~ {\rm Hz} ~ \laq ~\om ~\laq ~ 100~ {\rm GHz}.
\label{41}
\eeq
Future gravity wave detectors, able to reach this sensitivity level,
will directly test string theory and Planck scale physics.

As a final remark, I would like to answer a question that Guido
Pizzella asked me last year at CERN, during the First Meeting on 
{\sl ``Detection of high-frequency gravitational waves"}. The
question was: ``How sound is the prediction of such a high graviton
background"?

It is difficult to answer, and probably I am not the right person to
answer this question, but I would like to suggest an analogy. It
seems to me that we are in a situation similar, in some respect, to
the situation of many years ago in cosmology, when we had to
compare the steady-state model, and the hot-big bang model. One
of the main differences between the two models was just the
background of thermal radiation. Now, how sound was the
prediction of the thermal black body spectrum, before the
experimental discovery\cite{31} of Penzias and Wilson?

Difficult to say. The present situation seems to be similar. There are
standard inflationary models that predict a low background of
cosmic gravitons, at high frequency, and there are other models,
based on string theory, that predict a much higher background. How
sound are such predictions?

In my opinion, the experimentalist should tell us how sound are the
predictions, and not the converse. The answer should come, and
may come (in a not so far future), from experiments. Whatever the
final result may be, the experimental search for a cosmic graviton
background will become as important, for cosmology, as the study
of the electromagnetic CMB radiation. Even more important, in
some sense, because the thermal photons are relic radiation from
the big bang, while the cosmic graviton of a background like that of
eq. (\ref{41}) would be relic radiation from a much earlier pre-big
bang phase, preceding the hot, standard regime.  

\vspace{1cm}
{\it Acknowledgments:\/} I am grateful to  Ramy Brustein Massimo
Giovannini, Slava Mukhanov and Gabriele 
Veneziano for many useful discussions, and for a fruitful
collaboration. I wish to thank also the Organizing Committee for
their kind  invitation, and for the perfect organization of this
interesting Conference.

\end{document}